\documentclass[twocolumn]{aastex701}
\usepackage{amsmath}
\usepackage{booktabs}
\usepackage{array}
\usepackage{enumitem}
\newcommand{\Msun}{\mathrm{M}_{\odot}}
\definecolor{trackblue}{HTML}{1F77B4}
\definecolor{trackgreen}{HTML}{2CA02C}
\definecolor{trackorange}{HTML}{FF7F0E}
\definecolor{trackred}{HTML}{D62728}

\shorttitle{Diffuse and DM-deficient Tidal Remnants}
\shortauthors{He \& Jiang}

\begin{document}
\title{Tidal Evolution of Satellite Dwarf Galaxies. II. Ultra-diffuse and Dark-matter-deficient Remnants}
\author[orcid=0000-0003-0303-4188,gname=Feihong,sname=He]{Feihong He}
\affiliation{Kavli Institute for Astronomy and Astrophysics, Peking University, Beijing 100871, China}
\email[show]{feihong.he@pku.edu.cn}
\author[orcid=0000-0001-6115-0633,gname=Fangzhou,sname=Jiang]{Fangzhou Jiang}
\affiliation{Kavli Institute for Astronomy and Astrophysics, Peking University, Beijing 100871, China}
\email[show]{fangzhou.jiang@pku.edu.cn}
\correspondingauthor{Fangzhou Jiang}

\begin{abstract}
Dwarf galaxies exhibit substantial structural diversity, with some displaying both low surface brightness---characteristic of ultra-diffuse galaxies (UDGs), and a deficit of dark matter (DM) within their stellar extent. We examine these joint requirements with the suite of controlled simulations of dwarf satellites and the empirical stellar tidal tracks presented in Paper~I of this series. The suite probes the difference in tidal response of satellite structures upon variations in the initial subhalo concentration and inner density slope, as well as the stellar mass fraction, compactness, and S\'ersic index. We find that diffuseness and DM deficiency develop along opposite tidal pathways. Concentrated stellar distributions retain their stars as the halo is stripped and contract into dense, stellar-dominated remnants. Shallow stellar distributions expand or lose surface density, but they survive only while they retain a substantial DM contribution. Satellites with both low surface brightness and DM deficiency arise only within a narrow region of parameter space, characterised by a shallow inner subhalo density slope and an initial stellar distribution with a low S\'ersic index ($n_{\rm sersic}\leq1$). These tidally induced, DM-deficient UDGs survive for at most one or two apocentric passages before undergoing runaway disruption. We also use empirical tidal tracks to identify possible progenitors of NGC~1052-DF2 and NGC~1052-DF4. Considering both the availability of sufficiently massive stellar systems and the stripping needed to reach the observed masses and sizes, we favour progenitor halo masses near $10^{10.5}\,\Msun$ for both galaxies. 
\end{abstract}

\section{Introduction}\label{sec:intro}
In the standard cosmological picture, galaxies form in dark matter (DM) haloes, whose gravitational potentials allow gas to cool and form stars \citep{white1978,white1991}. Dwarf galaxies convert only a small fraction of their baryons into stars and are generally DM-dominated \citep{Bullock2017,Sales2022}. Their internal dynamics therefore probe the connection between galaxies and their haloes.

Dwarfs that appear to contain little or no DM are particularly informative. Such systems raise the question of how a galaxy can end up with a stellar body that is no longer embedded in a dominant halo. The best-studied examples are NGC~1052-DF2 and NGC~1052-DF4 (hereafter DF2 and DF4). Their low stellar velocity dispersions imply dynamical masses that can be largely accounted for by their stars alone \citep{vanDokkum2018a,Danieli2019,vanDokkum2019}. A similarly low DM fraction has been inferred for FCC~224 in the outskirts of the Fornax cluster \citep{Buzzo2025}, and follow-up spectroscopy of FCC~224 and its companion FCC~240 has enabled a direct comparison with the NGC~1052 pair \citep{Buzzo2026}. The puzzle posed by these galaxies is not only their DM deficiency: they also have extended, low-surface-brightness stellar components characteristic of ultra-diffuse galaxies (UDGs).

One proposed formation mechanism is a high-velocity collision between gas-rich dwarf galaxies. In this bullet-dwarf scenario, the collision shocks the gas and separates it from the collisionless DM. The gas subsequently collapses into stellar systems with little associated DM \citep{Silk2019,Shin2020}. The spatial and kinematic alignment of several diffuse galaxies in the NGC~1052 field led \citet{vanDokkum2022} to propose a common collisional origin for DF2 and DF4. Hydrodynamical simulations have since explored the formation of multiple DM-deficient galaxies along such a trail \citep{Lee2024}.

It has also been proposed that tidal stripping of dwarf satellites can produce DM-deficient satellites. After the dwarf falls into a more massive host, tides preferentially strip its extended halo, which can leave a remnant with a high stellar mass fraction. Controlled simulations have produced DF2-like remnants in this way, particularly from cored initial haloes \citep{Ogiya2018,Ogiya2022,Yang2020}. Cosmological simulations also identify DM-deficient satellites shaped by strong environmental interactions \citep{Moreno2022,He2026}. Tidal features detected around DF4 lend observational support to this mechanism \citep{Montes2020,Keim2022}. Relatedly, tidal heating can expand dwarfs and lower their surface brightness, offering a route to tidal UDGs \citep{Carleton2019,Jiang2019,Sales2020}.

In this picture, however, DM stripping is not a sufficient condition. Whether the stellar component survives substantial stripping of its DM envelope and becomes diffuse are key questions that previous studies have not adequately addressed. These two requirements pull in opposite directions. A centrally concentrated stellar distribution has enough self-gravity to resist further tidal disruption after losing its halo, but stripping tends to leave it compact and dense. A shallow, extended stellar distribution puffs up readily and loses surface density, but once the DM that shields it is removed, it is vulnerable to catastrophic disruption. Diffuseness and DM deficiency therefore impose competing survival requirements. Numerical case studies show that the two aspects can be reconciled, but they do not fully characterize which initial structures allow it, along which evolutionary paths, or for how long the joint state persists.

Addressing these questions requires following the stellar response of satellite galaxies across a broad range of progenitor structures. In \citet[][hereafter Paper~I]{PaperI}, we presented a library of 480 controlled simulations of dwarf satellites, spanning the halo concentration and inner slope as well as the stellar mass fraction, compactness, and S\'ersic index. We summarised the evolution of stellar mass, size, surface density, and velocity dispersion with empirical tidal tracks whose parameters depend on the initial structure. Here we apply that framework to study the formation of tidal UDGs and DM-deficient satellite dwarfs, and especially characterise the progenitor conditions for forming both diffuse and DM-deficient systems. We use it to map which regions of initial structural parameter space can produce DF2/DF4-like remnants, to trace the evolutionary trajectories that lead into and out of this regime, and to assess how long the simulated remnants reside there.

The paper is organised as follows. Section~\ref{sec:library} describes the simulation suite and the empirical stellar tidal tracks. Section~\ref{sec:results} defines the joint DM-deficient--ultra-diffuse selection, follows the remnants on two scaling diagrams, and uses the tracks to explain the competition between diffuseness and DM deficiency. Section~\ref{sec:discussion} considers the implications for formation scenarios (Section~\ref{sec:implications}) and uses the observed properties of DF2 and DF4 to constrain their possible tidal progenitors (Section~\ref{sec:progenitors}). Section~\ref{sec:summary} summarises our findings.

\section{Simulation and Tidal-track Library}\label{sec:library}
\subsection{Simulation Suite}\label{sec:suite}
We use the controlled simulations presented in Paper~I. Every satellite has an initial DM halo with a virial mass $m_{\rm vir}=10^{10}\,\Msun$ and evolves in a static, spherical NFW host \citep{NFW} with $M_{\rm vir}=10^{12}\,\Msun$ and concentration $c_{\rm h}=10$. Virial quantities are defined by a mean enclosed density of $200\rho_{\rm crit}$ at $z=0$, with $H_0=70\,\mathrm{km\,s^{-1}\,Mpc^{-1}}$, $\Omega_{\rm m}=0.3$, and $\Omega_\Lambda=0.7$. The satellite virial radius is $r_{\rm vir}=44.4\,\mathrm{kpc}$.

The satellite DM follows a smoothly truncated Dekel--Zhao profile \citep{Zhao1996,Dekel2017,Freundlich2020,Jiang2021}. Its concentration is $c_2=r_{\rm vir}/r_{-2}$, where $r_{-2}$ is the radius at which the logarithmic slope of the DM density $\rho_{\rm DM}$ is $-2$. We define $s_{001}=-{\rm d}\ln\rho_{\rm DM}/{\rm d}\ln r$ at $r=0.01r_{\rm vir}$, with $r$ the distance from the satellite centre. The stars follow a deprojected S\'ersic profile with index $n_{\rm sersic}$. We use $m_\star$ and $m_{\rm DM}$ to represent the bound stellar and DM masses, respectively, and $r_{\rm eff}$ for the projected stellar half-mass radius. A subscript 0 denotes an initial value. We vary the initial stellar-to-halo mass ratio $m_{\star,0}/m_{\rm vir}$, compactness $r_{\rm eff,0}/r_{\rm vir}$, and S\'ersic index independently of the halo parameters (Table~\ref{tab:grid}).

\begin{table}[!t]
\centering
\caption{Initial structural grid of the tidal-track library.\label{tab:grid}}
\small
\begin{tabular}{ll}
\toprule
Parameter & Values \\
\midrule
$m_{\rm vir}$ & $10^{10}\,\Msun$ \\
$c_2$ & $5,\ 10$ \\
$s_{001}$ & $0.6,\ 1.0,\ 1.2,\ 1.4,\ 1.8$ \\
$m_{\star,0}/m_{\rm vir}$ & $0.001,\ 0.01,\ 0.05$ \\
$r_{\rm eff,0}/r_{\rm vir}$ & $0.01,\ 0.02,\ 0.04,\ 0.1$ \\
$n_{\rm sersic}$ & $0.5,\ 1.0,\ 2.5,\ 4.0$ \\
\bottomrule
\end{tabular}
\par\smallskip
\begin{minipage}{\columnwidth}\footnotesize
\textit{Note.} The Cartesian product gives $2\times5\times3\times4\times4=480$ satellites. The DM halo mass inside $r_{\rm vir}$ is fixed. $r_{\rm eff,0}$ is the initial projected stellar half-mass radius. The index $n_{\rm sersic}$ labels the initial stellar profile.
\end{minipage}
\end{table}

Initial conditions are generated with \textsc{Agama} \citep{Vasiliev2019}, using isotropic distribution functions for both components in their combined potential. DM and stars have equal particle masses, $m_{\rm p}=10^4\,\Msun$, to suppress spurious mass-segregation heating \citep{Ludlow2019,Ludlow2021,Ludlow2023}. The halo initially contains $10^6$ particles inside $r_{\rm vir}$, and the gravitational softening is $\varepsilon=0.003r_{-2}$. We integrate each satellite for $20\,\mathrm{Gyr}$ with \textsc{Gadget-4} \citep{gadget4}. All initial orbits have an apocentre distance of 100 kpc, with pericentres between 5 and 50 kpc chosen for individual experiments to probe tidal stripping of different strength.

Snapshots are separated by $0.02\,\mathrm{Gyr}$. \textsc{HBT+} \citep{Han2012,Han2018} is used to identify the bound remnant, from which we measure the stellar and DM bound masses. The half-mass radius $r_{\rm eff}$ encloses half of the bound stellar mass projected perpendicular to the orbital plane. For a spatially constant stellar mass-to-light ratio, it equals the projected half-light radius. The stellar line-of-sight velocity dispersion is measured within the projected half-mass aperture for each of three orthogonal views and then averaged. 

A tidal track describes how a satellite's structural properties change as it loses bound mass \citep{Penarrubia2008}. For DM subhaloes, the maximum circular speed $v_{\rm max}$ and the radius $r_{\rm max}$ at which it is reached follow approximately the same tracks across different orbits for a given initial structure when measured at apocentre \citep{Errani2021}. Paper~I applied the same idea to the stellar component. We measured the stellar properties and bound mass at successive apocentres and used these measurements to fit the secular stellar response. Sampling at apocentre reduces the effect of oscillations due to tidal compression, heating and relaxation within each orbit. The fitted stellar tracks depend on the progenitor's halo and stellar structure.

\subsection{Tidal Tracks for Stellar Mass and Stellar Mass Fraction}\label{sec:mass}
The stellar tracks use the scaled total bound mass $x_{\rm bound}$ as their independent variable,
\begin{equation}
 x_{\rm bound}\equiv\frac{m_{\rm bound}}{m_{\rm max,0}},\qquad
 m_{\rm bound}=m_{\star}+m_{\rm DM},
 \label{eq:xbound}
\end{equation}
where $m_{\rm max,0}$ is the initial DM mass enclosed within the radius of the maximum circular velocity. The stellar mass track is
\begin{equation}
 f_{\star}\equiv\frac{m_{\star}}{m_{\star,0}}
 =1-\exp\!\left[-\ln2\left(\frac{x_{\rm bound}}{f_{1/2}}\right)^b\right].
 \label{eq:fstar}
\end{equation}
The half-mass stripping scale $f_{1/2}$ is the value of $x_{\rm bound}$ at which half of the initial stars remain bound, a measure of how much stripping the satellite undergoes before substantial stellar mass loss. The parameter $b$ controls the sharpness of the transition and sets the late-stage slope, $f_{\star}\propto x_{\rm bound}^{b}$.

The tidal-track parameters depend on the initial structure. Paper I quantified these dependencies empirically, including
\begin{align}
    \log_{10}f_{1/2} &\  \simeq  1.4 \log_{10}(r_{\rm eff,0}/r_{\rm vir}) +1,\label{eq:fhalf_scaling}\\
    b &\ \simeq 1.2\sqrt{\frac{s_{001}}{n_{\rm sersic}}}.\label{eq:b_scaling}
\end{align}
More compact stellar distributions remain shielded until a larger fraction of the halo has been stripped. The combination $\sqrt{s_{001}/n_{\rm sersic}}$ sets the sharpness, linking stellar retention to both the inner halo slope and the concentration of the stellar profile.

Define the global stellar-to-bound mass ratio as $\xi=m_{\star}/m_{\rm bound}$. Eq.~\ref{eq:fstar} gives
\begin{equation}
 \frac{\xi}{\xi_0}=\frac{x_{\rm bound,0}}{x_{\rm bound}}f_{\star}.
 \label{eq:xi}
\end{equation}
The ratio initially rises as the outer halo is removed while the stars remain intact. In the strongly stripped regime, $\xi/\xi_0\propto x_{\rm bound}^{b-1}$. For $b<1$, the stellar contribution continues to rise until it approaches the physical bound $\xi=1$. For $b>1$, the stellar mass falls faster than the total bound mass, and the ratio turns over. The aforementioned trend of the stellar mass ratio thus quantifies preferential stripping. To quantify DM deficiency in the simulations, we instead measure the stellar fraction inside $r_{\rm eff}$, as defined in Section~\ref{sec:selection}.

\subsection{Tidal Tracks for Stellar Size and Stellar Surface Density}\label{sec:size}
The empirical size track combines the response to stellar mass loss with an expansion factor
\begin{align}
 \frac{r_{\rm eff}}{r_{\rm eff,0}}&=P(x_{\rm bound})f_{\star}^{\beta},\label{eq:reff}\\
 P(x_{\rm bound})&\equiv1+A\exp\!\left[-\left(\frac{x_{\rm bound}}{f_{\rm p}}\right)^b\right].\label{eq:puff}
\end{align}
Here, $A$ sets the amplitude of the expansion term, $f_{\rm p}$ is a characteristic scaled mass, and $b$ is the same power-law index controlling the sharpness of transition as in the stellar mass track. The size-response index $\beta$ describes the change in radius as stellar mass is stripped. Once $P$ approaches its asymptotic value of $1+A$ in the deeply stripped regime, $\beta$ alone determines the size evolution: $\beta>0$ implies size decrease, while $\beta<0$ implies expansion. 

Here, the empirical dependence of the track parameter $\beta$ on the initial structures is approximated by
\begin{equation}
 \beta\simeq-0.77+0.66\sqrt{n_{\rm sersic}}+0.3s_{001}.
 \label{eq:beta_scaling}
\end{equation}
More concentrated stellar profiles therefore tend to contract more strongly. The impact of the inner halo slope on this response is secondary.


The mean projected stellar surface density is given by
\begin{equation}
 \Sigma_{\rm eff}=\frac{m_{\star}}{2\pi r_{\rm eff}^{2}},
 \label{eq:sigma_eff}
\end{equation}
so that the mass and size tracks together construct a tidal track of stellar surface density,
\begin{equation}
 \frac{\Sigma_{\rm eff}}{\Sigma_{\rm eff,0}}
 =P(x_{\rm bound})^{-2}f_{\star}^{1-2\beta}.
 \label{eq:sigma}
\end{equation}
The boundary $\beta=1/2$ separates secular evolution towards a lower surface density from evolution towards a denser remnant. This surface-density boundary differs from the boundary $\beta=0$ that separates contraction from expansion of the radius. A dwarf can therefore decrease in size while becoming more diffuse if its stellar mass falls faster than its projected area. Table~\ref{tab:tracks} summarises the tracks and their roles in driving the structural diversity of satellite dwarfs. 

\begin{table*}[!t]
\centering
\caption{Stellar tidal tracks that connect stripping to DM deficiency and diffuseness.\label{tab:tracks}}
\footnotesize
\begin{tabular}{>{\raggedright\arraybackslash}p{0.96in}p{2.65in}>{\raggedright\arraybackslash}p{2.72in}}
\toprule
Quantity & Functional form& Physical meanings of parameters\\
\midrule
Stellar mass & $f_{\star}=1-\exp[-\ln2\,(x_{\rm bound}/f_{1/2})^b]$ & $f_{1/2}$ sets the depth of stripping needed to lose half of the stars. $b$ sets the late mass-loss slope. \\[4pt]
Stellar-to-bound mass ratio & $\xi/\xi_0=(x_{\rm bound,0}/x_{\rm bound})f_{\star}$ & At late stages, $\xi\propto x_{\rm bound}^{b-1}$. The stellar contribution rises for $b<1$ and declines for $b>1$. \\[4pt]
Stellar size & $r_{\rm eff}/r_{\rm eff,0}=P(x_{\rm bound})f_{\star}^{\beta}$ & $A$ and $f_{\rm p}$ control puff-up. After it saturates, $\beta>0$ gives contraction and $\beta<0$ expansion. \\[4pt]
Mean surface density & $\Sigma_{\rm eff}/\Sigma_{\rm eff,0}=P(x_{\rm bound})^{-2}f_{\star}^{1-2\beta}$ & $\beta<1/2$ gives lower surface density and $\beta>1/2$ a denser remnant. \\[4pt]
\bottomrule
\end{tabular}
\par\smallskip
\begin{minipage}{\textwidth}\footnotesize
\textit{Note.} $P(x_{\rm bound})=1+A\exp[-(x_{\rm bound}/f_{\rm p})^b]$. The track parameters depend on initial structure, approximated by $f_{1/2}\propto(r_{\rm eff,0}/r_{\rm vir})^{1.4}$, $b\propto\sqrt{s_{001}/n_{\rm sersic}}$, and Eq.~\ref{eq:beta_scaling} for $\beta$. The surface density and stellar-to-bound mass ratio can be derived from the stellar mass and size tracks without introducing additional fitting parameters. 
\end{minipage}
\end{table*}

\section{Ultra-diffuse and DM-deficient Tidal Remnants}\label{sec:results}

\subsection{Joint Selection and Simulated Pathways}\label{sec:selection}

\begin{figure*}[!t]
\centering
\includegraphics[width=0.98\textwidth]{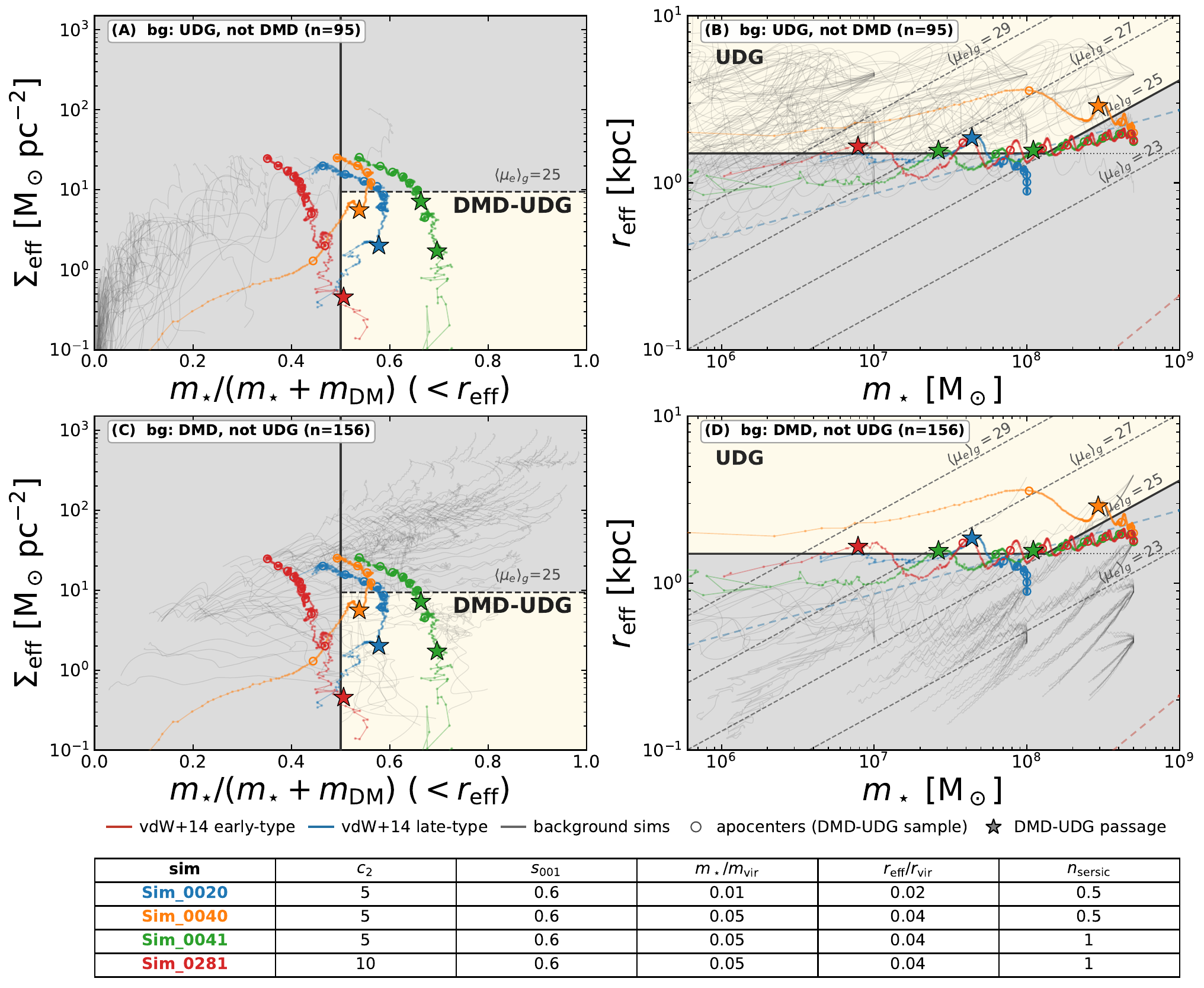}
\caption{Tidal evolution of the simulated satellites in observable spaces spaned by the stellar effective radius ($r_{\rm eff}$), stellar mass ($m_\star$), surface stellar density within effective radius ($\Sigma_{\rm eff}$), and stellar mass fraction within effective radius ($m_{\star}/(m_{\star}+m_{\rm DM})(<r_{\rm eff})$). The yellow shaded region marks DM-deficient UDGs (DMD-UDGs), defined by Eqs.~\ref{eq:dmd} and~\ref{eq:udg}. Coloured curves show the four simulations that enter this region at apocentre, and the table below the panels lists their initial structural parameters. Coloured circles mark their apocentre snapshots, and stars mark the snapshots that satisfy the DMD-UDG criteria. Grey curves in panels (A) and (B) show the 95 satellites that become ultra-diffuse but not DM-deficient. Grey curves in panels (C) and (D) show the 156 satellites that become DM-deficient but not ultra-diffuse. The left panels (A, C) show the mean stellar surface density within the half-mass radius, $\Sigma_{\rm eff}$, versus the stellar mass fraction within effective radius, $m_{\star}/(m_{\star}+m_{\rm DM})(<r_{\rm eff})$. The right panels (B, D) show the projected half-mass radius versus stellar mass. Diagonal lines mark constant mean surface brightness. 
The blue and red dashed lines show the size--mass relations of late- and early-type galaxies from \citet{vanderwel2014}. Satellites that become ultra-diffuse but not DM-deficient evolve towards lower $\Sigma_{\rm eff}$ and lower $m_{\star}/(m_{\star}+m_{\rm DM})(<r_{\rm eff})$, whereas satellites that become DM-deficient but not ultra-diffuse contract towards higher $\Sigma_{\rm eff}$. All four DMD-UDG progenitors have the shallowest inner halo slope in the grid ($s_{001}=0.6$) and $n_{\rm sersic}\leq1$. Each remains in the joint region for only one or two apocentres.}
\label{fig:sample}
\end{figure*}

We now use the tidal-track simulation suite to examine whether DM-deficient UDGs can form through tidal stripping and which initial halo and stellar structures favour their formation. We test both conditions at the same apocentre. For DM deficiency we require that stars contribute more than half of the mass inside the effective radius,
\begin{equation}
 F_{\star,\rm e}\equiv
 \frac{m_\star(<r_{\rm eff})}
      {m_\star(<r_{\rm eff})+m_{\rm DM}(<r_{\rm eff})}
 >\frac{1}{2}.
 \label{eq:dmd}
\end{equation}
Here the enclosed masses are evaluated within a sphere of radius $r_{\rm eff}$, whereas $r_{\rm eff}$ itself is measured in projection. This enclosed criterion differs from the global stellar-to-bound mass ratio $\xi$ defined in Eq.~\ref{eq:xi}.

We apply the DM-deficiency criterion in Eq.~\ref{eq:dmd} to the simulated remnants. For the UDG selection, we use the mean effective stellar surface density defined in Eq.~\ref{eq:sigma_eff}, adopting a constant $g$-band stellar mass-to-light ratio of $(M/L)_g=2\,\Msun/L_{\odot,g}$. We require
\begin{equation}
r_{\rm eff}\geq1.5\,{\rm kpc},
\qquad
\Sigma_{\rm eff}\leq9.415\,\Msun\,{\rm pc}^{-2},
\label{eq:udg}
\end{equation}
where the surface-density threshold corresponds to a mean effective surface brightness of $\langle\mu_{\rm e}\rangle_g\geq25\,{\rm mag\,arcsec^{-2}}$. Systems satisfying both the DM-deficiency and UDG criteria are referred to as DM-deficient UDGs (DMD-UDGs).

\begin{figure*}[!t]
\centering
\includegraphics[width=\textwidth]{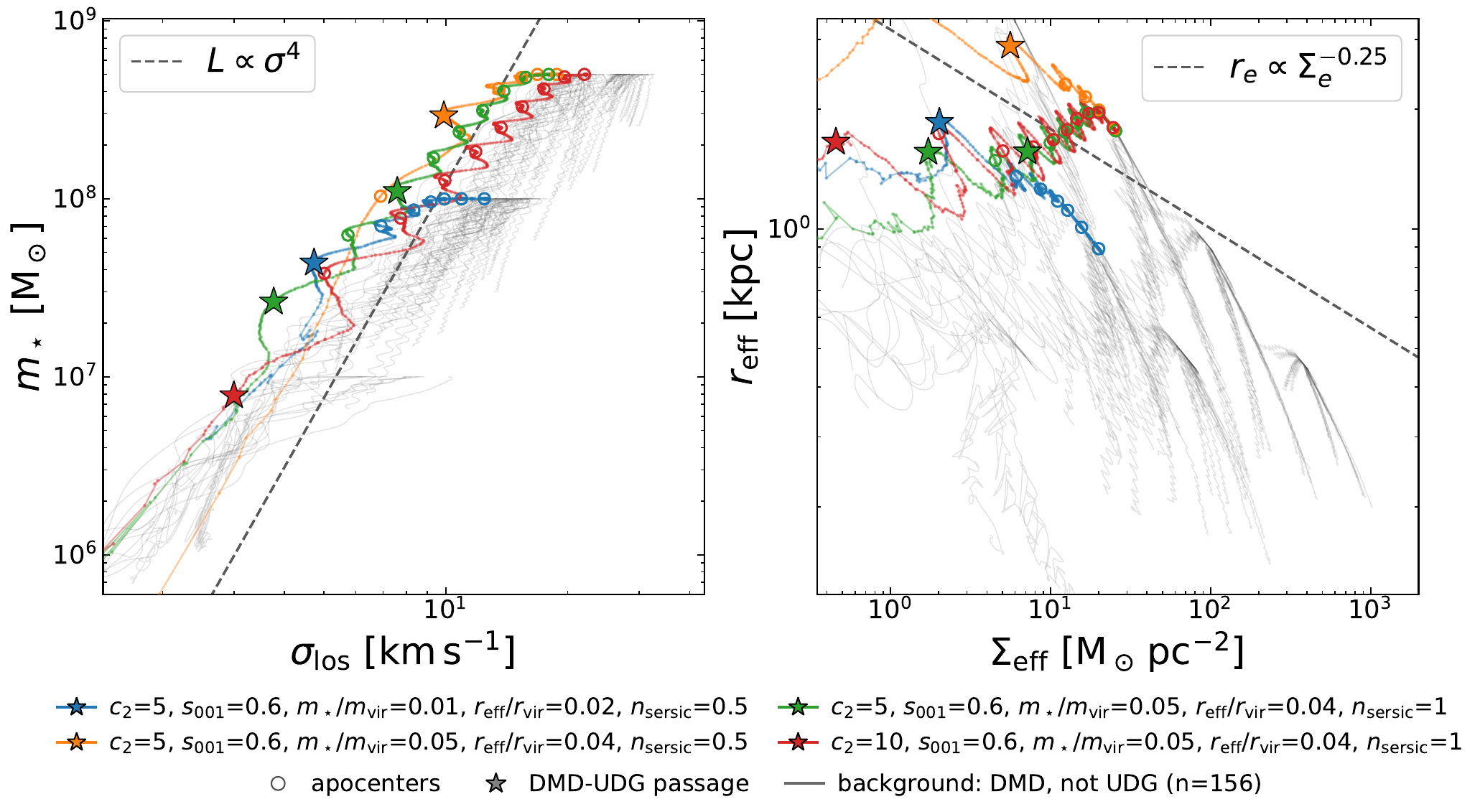}
\caption{Tidal tracks of DM-deficient remnants in the spaces of galaxy scaling relations: the plane of stellar mass and stellar line-of-sight velocity dispersion $\sigma_{\rm los}$ within the projected half-mass radius (left), and the plane of projected half-mass radius $r_{\rm eff}$ and mean stellar surface density $\Sigma_{\rm eff}$ (right). Grey curves show the 156 simulated satellites that become DM-deficient but not ultra-diffuse. Coloured curves show the four DMD-UDG candidates, whose initial parameters are listed in the legend and in Fig.~\ref{fig:sample}. Circles mark their apocentres, and stars mark snapshots that satisfy the joint DMD-UDG selection. The dashed lines show the classical Faber--Jackson ($L\propto\sigma^4$) and Kormendy relations for reference. For the Faber--Jackson relation, stellar mass serves as a proxy for luminosity at a fixed mass-to-light ratio. As DM is stripped, the DM-deficient, non-UDG remnants lose stellar mass and velocity dispersion roughly parallel to the Faber--Jackson relation, and most contract towards higher $\Sigma_{\rm eff}$. The DMD-UDG candidates instead reach low velocity dispersions together with much lower surface densities.}
\label{fig:scaling_dmd}
\end{figure*}

\begin{figure*}[!t]
\centering
\includegraphics[width=\textwidth]{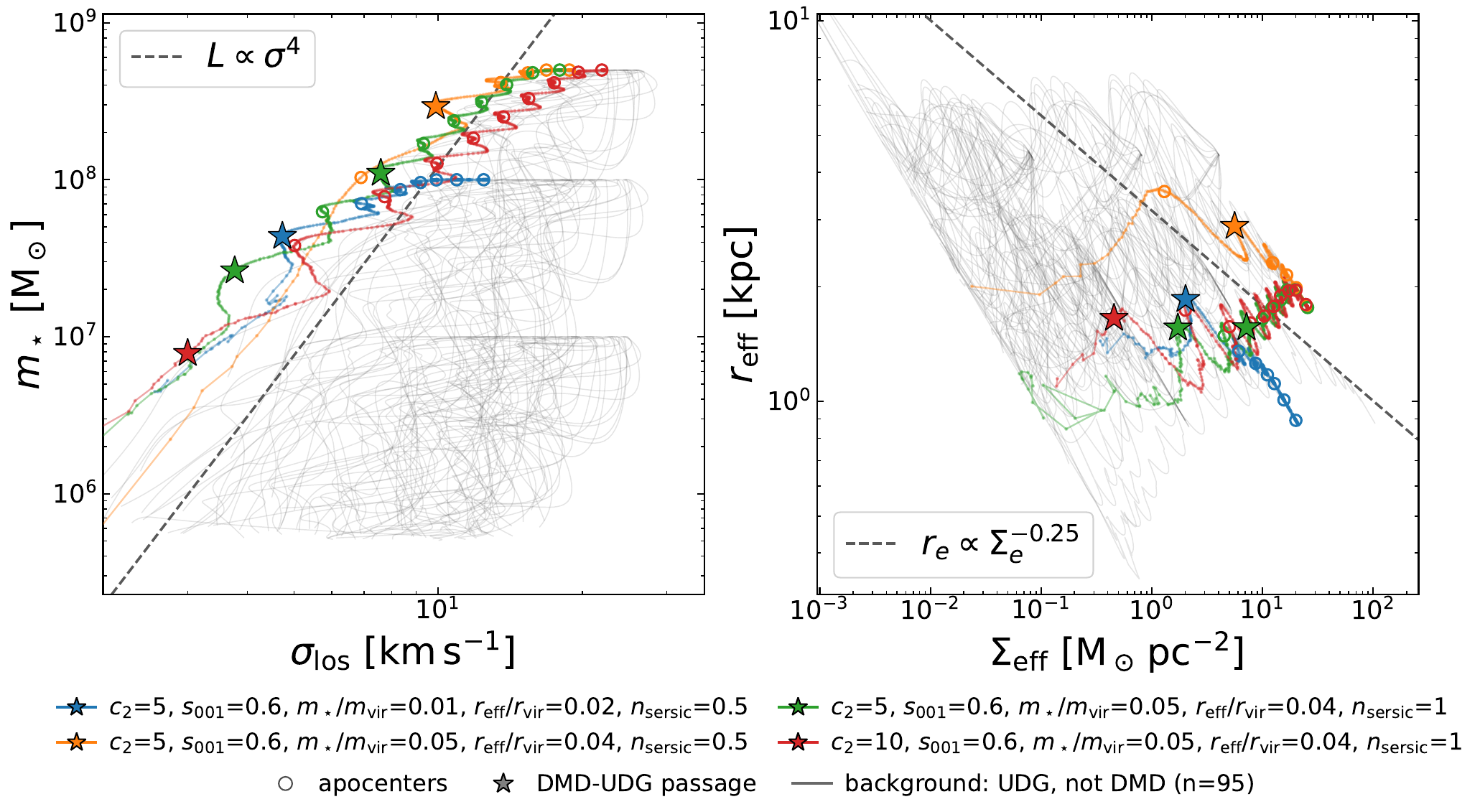}
\caption{Same as Fig.~\ref{fig:scaling_dmd}, but with grey curves showing the 95 simulated satellites that become ultra-diffuse without becoming DM-deficient. The coloured curves, symbols, and dashed reference lines are as in Fig.~\ref{fig:scaling_dmd}. These UDG remnants remain DM-supported. At fixed stellar mass, they retain larger velocity dispersions than the DM-deficient remnants in Fig.~\ref{fig:scaling_dmd}, and they evolve towards lower $\Sigma_{\rm eff}$, often while expanding. The four DMD-UDG candidates follow this diffuse population in the size--surface-density plane but reach much lower velocity dispersions. They therefore occupy the narrow overlap between the diffuse, DM-dominated population shown here and the compact, stellar-dominated population in Fig.~\ref{fig:scaling_dmd}.}
\label{fig:scaling_udg}
\end{figure*}

We evaluate these criteria at apocentre passages, where the stellar structure is close to equilibrium. Only four of the 480 simulated dwarfs have at least one apocentre snapshot within the DMD-UDG region, as shown in Fig.~\ref{fig:sample}. All four originate from haloes with the shallowest inner slope in the grid ($s_{001}=0.6$) and have low S\'ersic indices ($n_{\rm sersic}\leq1$). These initial structures favour both substantial DM stripping and a reduction in stellar surface density, which allows the remnants to enter the DMD-UDG region. The same structures, however, make the systems vulnerable to further tidal stripping. All four candidates subsequently undergo runaway disruption. Each spends only one or two apocentre passages in the DMD-UDG region before losing its bound structure. The DMD-UDG phase is therefore a short-lived stage preceding disruption rather than a stable evolutionary endpoint. If observed DMD-UDGs formed in this way, they would be transient systems caught shortly before disruption.

Considering the two criteria separately clarifies this tension. In the upper panels of Fig.~\ref{fig:sample}, satellites that become ultra-diffuse but remain DM-dominated move towards lower $\Sigma_{\rm eff}$ as their stellar fraction within $r_{\rm eff}$ decreases. The two DMD-UDG cases with $n_{\rm sersic}=0.5$ reflect a two-stage path. Their stellar fraction initially rises as the outer DM is removed, reaches a maximum, and then declines once stellar stripping becomes efficient. One candidate leaves the DM-deficient region by the following apocentre, which directly illustrates the transient nature of the DMD-UDG phase. By contrast, the vast majority of the DM-deficient remnants never become ultra-diffuse. Their half-mass radii contract and their surface densities increase. They therefore evolve towards compact, stellar-dominated remnants rather than towards UDGs.


Our parameter scan therefore supports a more restrictive interpretation of the tidal scenario. Consistent with previous controlled studies, tides can form DF2/DF4-like DMD-UDGs; however, the progenitors that do so occupy a small region of structural parameter space, and the resulting phase is short-lived. A successful match along a single idealised trajectory thus reflects a specific combination of progenitor structure, orbit, and observation time rather than a common outcome of tidal evolution. Establishing the true occurrence rate requires large samples from cosmological hydrodynamical simulations that resolve the internal stellar and DM structure of satellites well enough to follow the strongly stripped regime.

\subsection{Tidal Tracks on Classical Scaling Diagrams}\label{sec:scaling}

We next follow the DM-deficient and UDG-like remnants identified in Section~\ref{sec:selection} on two classical scaling diagrams, the stellar mass--velocity dispersion plane and the size--surface-density plane (Figs.~\ref{fig:scaling_dmd} and \ref{fig:scaling_udg}). For reference, we include the Faber--Jackson relation ($L\propto\sigma^4$; \citealt{Faber1976}) and the Kormendy relation \citep{Kormendy1977}, both originally established for luminous early-type galaxies. Observed dwarf galaxies deviate from simple extrapolations of these relations. They follow a shallower Faber--Jackson relation and different size--surface-brightness trends \citep{Toloba2012,Kourkchi2012a,Kourkchi2012b,Boselli2008}.

The left panels of Figs.~\ref{fig:scaling_dmd} and \ref{fig:scaling_udg} show the tracks in the $m_{\star}$--$\sigma_{\rm los}$ plane, where stellar mass serves as a proxy for luminosity at a fixed mass-to-light ratio. As tidal stripping proceeds, both stellar mass and velocity dispersion decrease. The DM-deficient remnants in Fig.~\ref{fig:scaling_dmd} evolve broadly along the direction of the classical Faber--Jackson relation, and some tracks run approximately parallel to the $L\propto\sigma^4$ scaling over part of their evolution. This behaviour is consistent with the growing role of stellar self-gravity in setting the central velocity dispersion as DM is removed, as in the stellar-dominated inner regions of luminous elliptical galaxies. In contrast, the tracks of the UDG remnants in Fig.~\ref{fig:scaling_udg} are more widely spread. These remnants often retain relatively large velocity dispersions at a given stellar mass, consistent with their continued DM support. At late stages, the four DMD-UDG candidates reach much lower velocity dispersions, approaching the values inferred for DF2 and DF4 ($\lesssim 10\,{\rm km\,s^{-1}}$).

The right panels of Figs.~\ref{fig:scaling_dmd} and \ref{fig:scaling_udg} show the corresponding tracks in the $r_{\rm eff}$--$\Sigma_{\rm eff}$ plane. Here, the two populations respond in opposite ways. Most DM-deficient remnants contract and develop higher stellar surface densities, whereas UDG remnants evolve towards lower surface densities, often while growing in size. The DMD-UDG candidates follow the latter trend towards low surface density, although, depending on the candidate, their radii increase, remain nearly unchanged, or decrease moderately during stripping. Their trajectories therefore do not follow a single Kormendy-like scaling. Instead, the evolution in this plane reflects the competition between stellar mass loss and the structural response of the remaining stars.



\subsection{Toy Model for the Tidal Responses}\label{sec:competition}


\begin{figure*}[!t]
\centering
\includegraphics[width=0.47\textwidth]{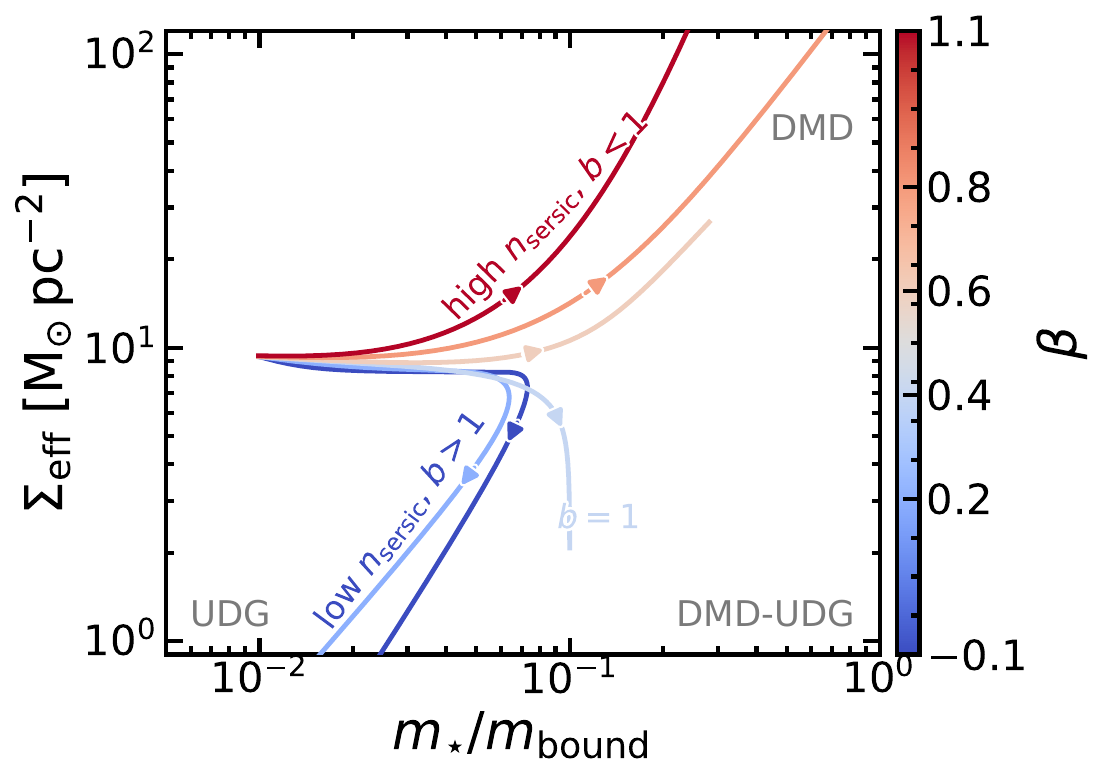}\hfill
\includegraphics[width=0.51\textwidth]{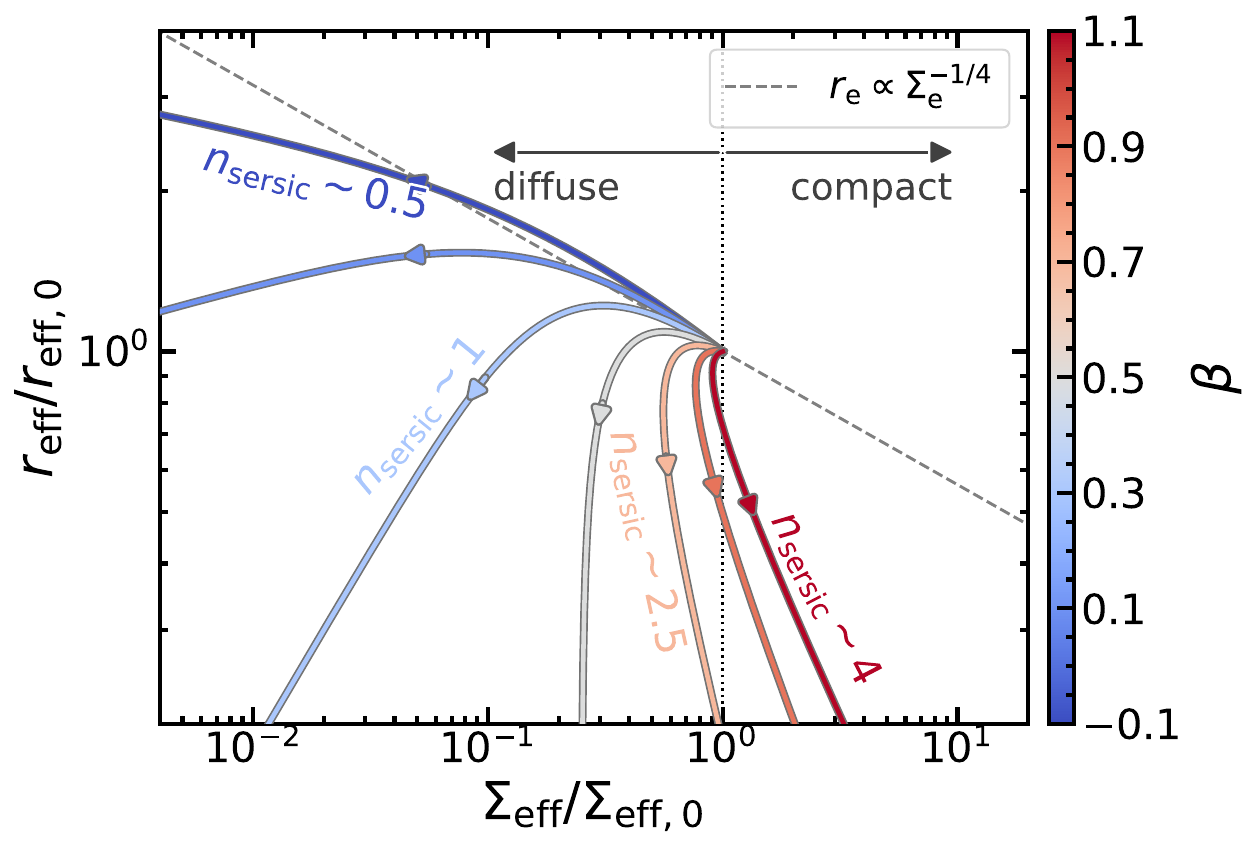}
\caption{Tidal evolution predicted by the empirical stellar tidal tracks (Table~\ref{tab:tracks}), illustrating why DM deficiency and diffuseness are usually reached along different pathways. Arrows indicate increasing tidal mass loss. \textit{Left:} mean stellar surface density $\Sigma_{\rm eff}$ versus the global stellar-to-bound mass ratio $\xi=m_{\star}/m_{\rm bound}$. The stellar mass-loss sharpness $b$ is varied together with $\beta$ to mimic their joint dependence on $n_{\rm sersic}$, with $A=1$, $f_{1/2}=f_{\rm p}=0.1$, and initial values $\xi=10^{-2}$ and $\Sigma_{\rm eff}=10\,\Msun\,\mathrm{pc}^{-2}$. Tracks with $b<1$ and $\beta>1/2$, typical of high-$n_{\rm sersic}$ systems, evolve towards higher surface density and stellar dominance. Tracks with $b>1$ and $\beta<1/2$, typical of low-$n_{\rm sersic}$ systems, evolve towards lower surface density, and their stellar-to-bound mass ratio eventually turns over. \textit{Right:} normalised radius $r_{\rm eff}/r_{\rm eff,0}$ versus normalised surface density $\Sigma_{\rm eff}/\Sigma_{\rm eff,0}$, with $A=b=1$ and $f_{1/2}=f_{\rm p}=0.01$ fixed and only $\beta$ varied. Tracks with $\beta<1/2$, characteristic of low-$n_{\rm sersic}$ systems, move towards lower surface density and can puff up. Tracks with $\beta>1/2$ contract towards higher surface density. The grey dashed line shows the Kormendy relation for luminous elliptical galaxies for reference.}
\label{fig:illustration}
\end{figure*}



The empirical stellar tidal tracks of Paper~I, summarised in Section~\ref{sec:library} and Table~\ref{tab:tracks}, explain the contrasting tidal pathways identified above. Combining the stellar mass and size tracks gives the evolution of the mean stellar surface density. The stellar mass track and the bound mass together determine the global stellar-to-bound mass ratio, $\xi=m_{\star}/m_{\rm bound}$. Fig.~\ref{fig:illustration} shows these relations in two projections, the $\Sigma_{\rm eff}$--$\xi$ plane and the $r_{\rm eff}$--$\Sigma_{\rm eff}$ plane. We fix different parameters in the two panels to isolate the effects of the stellar mass-loss sharpness $b$ and the size-response index $\beta$.

The left panel shows how stellar concentration shapes the competing evolution of surface density and stellar mass fraction. We set $A=1$ and $f_{1/2}=f_{\rm p}=0.1$, with initial values $\xi=10^{-2}$ and $\Sigma_{\rm eff}=10\,\Msun\,{\rm pc}^{-2}$. We vary $b$ and $\beta$ together to mimic their joint dependence on the initial S\'ersic index $n_{\rm sersic}$. Concentrated stellar distributions generally have smaller $b$ and larger $\beta$. For $b<1$ and $\beta>1/2$, the stellar-to-bound mass ratio continues to rise as DM is stripped, while the stellar body contracts enough to increase its surface density. In contrast, shallow stellar distributions tend to have $b>1$ and $\beta<1/2$. Their surface density declines, and their stellar-to-bound mass ratio eventually turns over as stellar stripping becomes efficient. These trends explain why DM deficiency and diffuseness are usually reached along different tidal pathways.

In the right panel, we isolate the size response by fixing $A=b=1$ and $f_{1/2}=f_{\rm p}=0.01$ and varying only $\beta$. From the empirical tracks, $r_{\rm eff}/r_{\rm eff,0}=P f_{\star}^{\beta}$ and $\Sigma_{\rm eff}/\Sigma_{\rm eff,0}=P^{-2}f_{\star}^{1-2\beta}$. Once the puff-up factor $P$ has saturated, the critical value separating increasing and decreasing surface densities is $\beta=1/2$. For $\beta<1/2$, stellar mass loss outweighs the change in projected area, so the surface density decreases even if the stellar radius contracts moderately. For $\beta>1/2$, the radius contracts rapidly enough that the surface density increases despite stellar mass loss. Different values of $\beta$ therefore produce distinct evolutionary directions in the size--surface-density plane rather than a single track along the classical Kormendy relation. Because $\beta$ depends strongly on the initial S\'ersic index and additionally on the inner DM slope, these responses reflect the diversity of progenitor structures.

The empirical size track describes the secular evolution sampled at apocentre rather than the full response along the orbit. In the simulations, the stellar radius oscillates substantially over each orbit. The oscillations are strongest for initially shallow stellar distributions with low S\'ersic indices, whose stars respond more strongly to tidal perturbations. These variations in $r_{\rm eff}$ also produce fluctuations in $\Sigma_{\rm eff}\propto m_\star/r_{\rm eff}^2$, which give rise to the oscillatory trajectories in Figs.~\ref{fig:scaling_dmd} and \ref{fig:scaling_udg}. The toy model thus captures the overall direction of tidal evolution, whereas the detailed trajectories also depend on orbital phase. In particular, the boundary $\beta=1/2$ characterises the secular surface-density response after the puff-up factor has saturated, not the instantaneous direction of evolution at every orbital phase.

In this illustrative model, we use the global stellar-to-bound mass ratio $\xi$ as a proxy for DM deficiency instead of the enclosed stellar fraction $F_{\star,\rm e}$ used to select DM-deficient remnants in the simulations. The global ratio follows directly from the empirical mass tracks, whereas the enclosed fraction requires additional information about the mass distribution. An approximate enclosed dynamical mass could be inferred by combining the velocity-dispersion and size tracks with a dynamical mass estimator \citep[e.g.,][]{Wolf2010}. However, our empirical velocity-dispersion track captures the dependence on the initial structural parameters less accurately than the mass and size tracks do. We therefore restrict this illustration to the global mass ratio.

The physical interpretation is that a DM-deficient remnant must retain enough stellar self-gravity to resist subsequent tides. Concentrated stellar components survive the loss of their haloes more readily but tend to become dense and compact. Diffuse stellar components have less self-gravity and require more DM to remain bound. Simultaneously obtaining low surface density and low DM content therefore requires a balance between these competing responses. The four DMD-UDG candidates in our simulations occupy this narrow overlap, and their subsequent disruption shows that such a configuration is difficult to maintain.

\section{Discussion}\label{sec:discussion}

\subsection{Implications for Formation Scenarios}\label{sec:implications}
Previous controlled simulations have shown that tidal stripping can produce DF2/DF4-like galaxies under favourable conditions, particularly for progenitors with shallow DM profiles \citep{Ogiya2018,Ogiya2022}. Our systematic survey identifies the structural constraints on this formation channel. The progenitors of DMD-UDGs occupy a narrow region of the initial parameter space, between progenitors that evolve into compact, DM-deficient remnants and progenitors that remain diffuse but DM-dominated. The shallow DM and stellar distributions required to satisfy both conditions also leave the remnants vulnerable to further stripping. All four DMD-UDG candidates subsequently undergo runaway disruption, with only one or two apocentre snapshots satisfying the joint selection. If DF2 and DF4 formed through tidal stripping, their observed properties would therefore represent a transient phase shortly before disruption rather than a long-lived evolutionary endpoint. Their abundance would then depend not only on the occurrence of suitable progenitors and orbits but also on the duration of this observable phase.



\subsection{Progenitors of DF2 and DF4}\label{sec:progenitors}

\begin{figure*}
    \centering
    \includegraphics[width=0.9\linewidth]{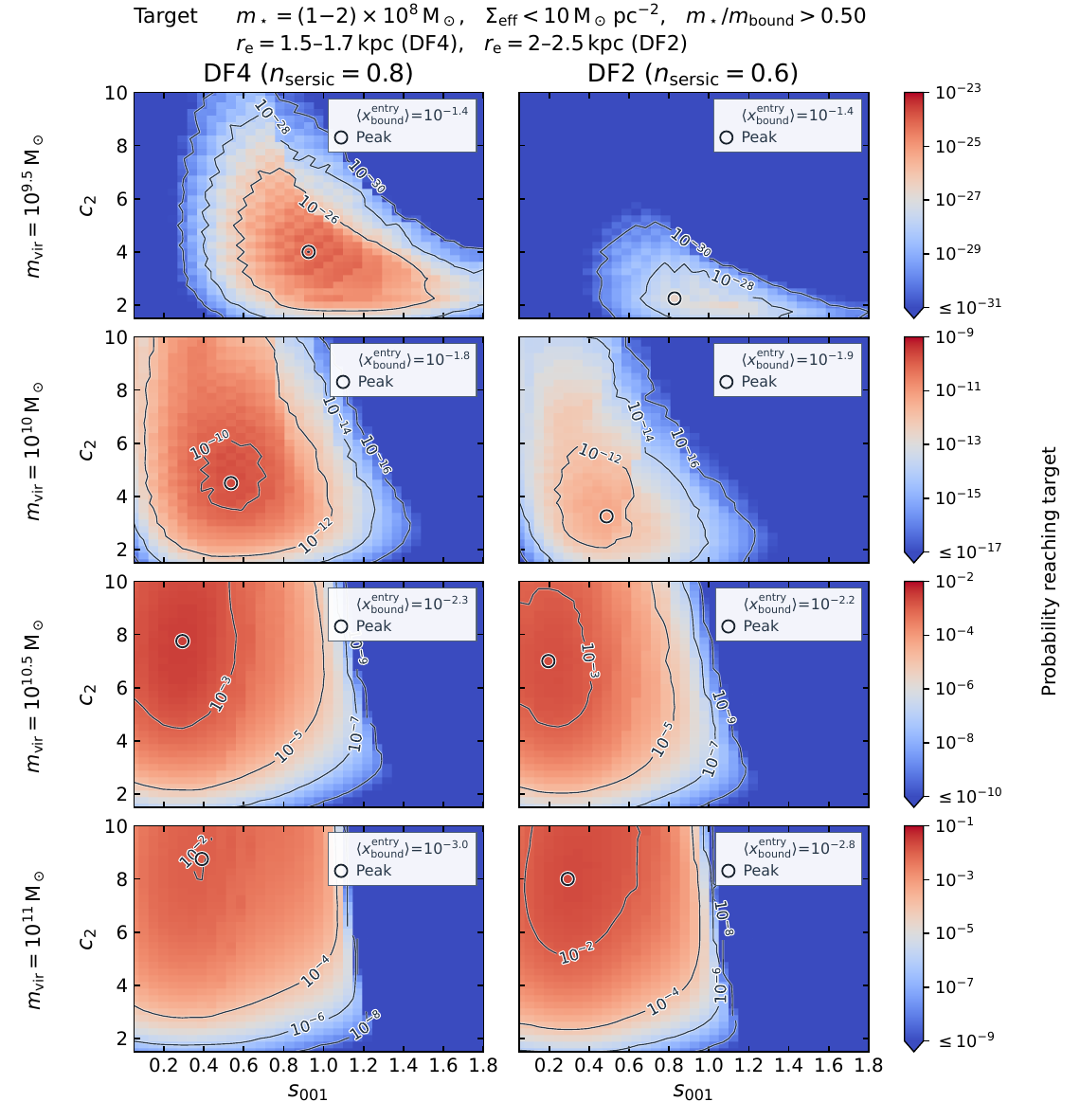}
    \caption{The probability of producing DM-deficient UDGs resembling DF2 and DF4 by tides, as a function of progenitor halo mass and structure. Rows correspond to different progenitor halo masses $m_{\rm vir}$, increasing from top to bottom. Colour shading and contours show the prior-weighted entry probability density per unit $c_2$ and $s_{001}$, marginalised over the initial stellar mass $m_{\star,0}$ and effective radius $r_{\rm eff,0}$. Each row uses its own logarithmic colour scale, with the darkest blue shading indicating regions of zero probability for forming a DF2/DF4-like tidal remnant. Open circles mark the probability maxima in each panel, and $\langle x_{\rm bound}^{\rm entry}\rangle$ denotes the weighted median scaled bound mass ($x_{\rm bound} \equiv m_{\rm bound}/m_{\rm max,0}$) when the tracks first enter the target window.}
    \label{fig:halo_condition_mass_scan}
\end{figure*}


In Section~\ref{sec:selection}, our simulation suite yielded only four DMD-UDG candidates within the explored parameter space. Although we have run 480 simulations, the multidimensional parameter space of satellite tidal evolution remains coarsely sampled. Because purely gravitational $N$-body evolution is scale-free, the dimensionless empirical tidal tracks established in Paper~I (Section~\ref{sec:library}), which depend only on the dimensionless structural ratios $m_{\star,0}/m_{\rm vir}$, $r_{\rm eff,0}/r_{\rm vir}$, $c_2$, $s_{001}$, and $n_{\rm sersic}$, can be generalised across different progenitor halo masses $m_{\rm vir}$. We therefore use this empirical model to extrapolate beyond the simulation grid and explore the preferred progenitor structural parameter space for DF2/DF4-like tidal remnants more broadly. 

To be classified as a viable candidate, a tidal remnant must satisfy the DMD-UDG selection criteria defined in Eqs.~\ref{eq:dmd} and~\ref{eq:udg}. As in Section~\ref{sec:competition}, because the empirical model does not track the internal density profile, we use the global stellar-to-bound mass ratio (Eq.~\ref{eq:xi} and Table~\ref{tab:tracks}) as a proxy for DM deficiency, requiring $\xi \equiv m_{\star}/m_{\rm bound} > 0.5$ along the track. 
The evolution of the stellar effective radius ($r_{\rm eff}$) and effective surface density ($\Sigma_{\rm eff}$) is computed via Eqs.~\ref{eq:reff} and~\ref{eq:sigma_eff} in Section~\ref{sec:library}, respectively. As discussed in Paper~I, the parameters $A$ and $f_{\rm p}$ are weakly constrained by the initial galaxy structure and remain difficult to model directly. We therefore adopt fixed fiducial values of $A = 0.4$ and $f_{\rm p} = 0.1$, corresponding to the medians from our simulation suite for $n_{\rm sersic} = 0.5$ and $1$. Meanwhile, $f_{1/2}$, $b$, and $\beta$ are modelled explicitly as functions of the initial dwarf structure. To match the observational constraints of DF2 (DF4), we require that a tidal track pass through the target window of $r_{\rm eff} = 2\text{--}2.5\,{\rm kpc}$ for DF2 ($1.5\text{--}1.7\,{\rm kpc}$ for DF4) when its stellar mass lies within $m_{\star} = (1\text{--}2) \times 10^{8}\,{\rm M}_{\odot}$ \citep{vanDokkum2018a, vanDokkum2019}, while simultaneously satisfying the remaining DMD-UDG criteria.

To evaluate the likelihood of different progenitor halo conditions under a tidal-stripping origin, we construct cosmologically motivated priors for the initial structural parameters across four fixed progenitor halo masses: $m_{\rm vir} = 10^{9.5}, 10^{10}, 10^{10.5},$ and $10^{11}\,{\rm M}_{\odot}$. We fix the initial S\'{e}rsic index to $n_{\rm sersic} = 0.6$ for DF2 ($0.8$ for DF4) \citep{vanDokkum2018a, vanDokkum2019}. Although $n_{\rm sersic}$ may in principle evolve during tidal stripping, our empirical model does not track $n_{\rm sersic}$ or the detailed evolution of the stellar density profile. Since the observed $n_{\rm sersic}$ values of DF2 and DF4 remain close to unity and vary only weakly, treating $n_{\rm sersic}$ as fixed is a reasonable approximation for our purposes. At each fixed $m_{\rm vir}$, we take empirical scaling relations as the median values and apply log-normal distributions around these medians to form the prior distributions. Specifically, the median initial stellar mass $m_{\star,0}$ follows the stellar-to-halo mass relation (SHMR) measured from the peak masses of satellites in the COLIBRE simulations \citep{He2026}. The halo concentration $c_{2}$ and inner density slope $s_{001}$ are then assigned following \textsc{SatGen} \citep{Jiang2021}. The dark-matter-only baseline concentration $c_{2,{\rm DMO}}$ is obtained from the mass--concentration relation of \citet{Zhao2009} and modified by the baryonic halo-response correction $(c_{2}/c_{2,{\rm DMO}})$ as a function of $(m_{\star,0}/m_{\rm vir})$. The median inner slope $s_{001}$ is determined by the NIHAO halo-response emulator \citep{NIHAO, Freundlich2020} as a function of the initial stellar-to-halo mass fraction. Finally, the median initial effective radius $r_{\rm eff,0}$ is tied to the progenitor halo virial radius and concentration via $r_{\rm eff,0}/r_{\rm vir} = 0.015\,(c_{2}/10)^{-0.7}$ \citep{Jiang2019size, Liang2025}\footnote{\citet{Jiang2019size} give this relation for the 3D stellar half-mass radius $r_{1/2}$ with a normalisation of $0.02$. Using the projected effective radius $r_{\rm eff}$ introduces a factor of $3/4$, yielding $0.015$.}, also with log-normal scatter.

Fig.~\ref{fig:halo_condition_mass_scan} presents the results of this tidal-track analysis. The colour scale in each panel shows the prior-weighted entry probability density per unit $c_2$ and $s_{001}$, marginalised over the initial stellar mass $m_{\star,0}$ and effective radius $r_{\rm eff,0}$, for tracks that enter the DF2 or DF4 target window at any point over $x_{\rm bound} \in [10^{-4}, 1]$. Redder regions denote higher prior-weighted formation probability densities (i.e., more favourable progenitor halo conditions), whereas dark blue regions indicate halo configurations that cannot produce a DF2/DF4-like remnant via tidal stripping. 

\begin{figure*}
    \centering
    \includegraphics[width=\linewidth]{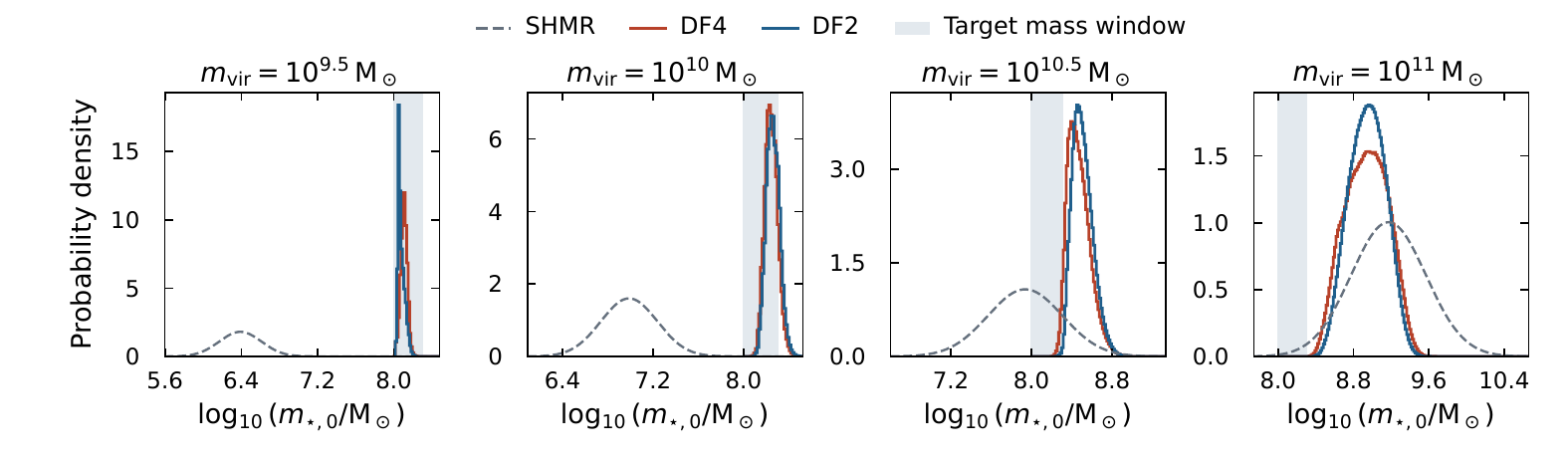}
    \caption{Initial stellar masses of progenitors whose tidal tracks reach the DF2 (blue) or DF4 (red) target. Panels run from $10^{9.5}$ to $10^{11}\,\Msun$ in progenitor halo mass. The dashed curve gives the stellar-mass prior at the halo mass shown, based on the SHMR extracted from the COLIBRE simulation \citep{He2026}. The coloured steps give the distributions conditional on reaching each target. The shaded band marks the common target interval in remnant stellar mass.}
    \label{fig:progneitor_stellar_mass}
\end{figure*}

Focusing first on the $m_{\rm vir} = 10^{10}\,{\rm M}_{\odot}$ case adopted in our simulation suite, we find that progenitor haloes with relatively low concentrations ($c_{2} \sim 3\text{--}4$) and cored inner profiles ($s_{001} \sim 0.5$) can potentially form DF2/DF4-like tidal remnants. However, cuspy haloes are ruled out as viable progenitors, and the model strongly favours haloes with $s_{001} \lesssim 1$. 

Furthermore, because $m_{\star,0}$ is weighted by the SHMR prior, a stellar mass of $m_{\star} \sim 10^{8}\,{\rm M}_{\odot}$ already lies $\sim 1\,{\rm dex}$ above the median relation for a $10^{10}\,{\rm M}_{\odot}$ halo, naturally suppressing the peak probability density (down to $\sim 10^{-10}$) in this panel. This suppression becomes even more pronounced at $m_{\rm vir} = 10^{9.5}\,{\rm M}_{\odot}$. Conversely, at higher progenitor masses, the peak probability density increases to $\sim 10^{-3}\text{--}10^{-2}$ at $m_{\rm vir} = 10^{10.5}\,{\rm M}_{\odot}$ and exceeds $10^{-2}$ at $m_{\rm vir} = 10^{11}\,{\rm M}_{\odot}$, as more massive haloes much more readily host dwarfs of the requisite stellar mass.

In Fig.~\ref{fig:progneitor_stellar_mass}, we show the probability distribution of the initial stellar mass ($m_{\star,0}$) for tracks that enter the DF2/DF4 target window at a given halo mass. For $m_{\rm vir} = 10^{10.5}\,{\rm M}_{\odot}$, the distribution peaks at $m_{\star,0} \simeq 3 \times 10^{8}\,{\rm M}_{\odot}$. For $m_{\rm vir} = 10^{11}\,{\rm M}_{\odot}$, progenitors with initial stellar masses as high as $m_{\star,0} \sim 10^{9}\,{\rm M}_{\odot}$ can be stripped down to remnants of $m_{\star} \sim 10^{8}\,{\rm M}_{\odot}$ via heavier tidal mass loss.

On the other hand, more massive progenitors require substantially stronger tidal stripping to enter the DF2/DF4 target mass window. Specifically, the median scaled bound mass upon entering the target region is $x_{\rm bound} = 10^{-2.2}$ for DF2 and $10^{-2.3}$ for DF4 at $m_{\rm vir} = 10^{10.5}\,{\rm M}_{\odot}$, compared to $10^{-2.8}$ (DF2) and $10^{-3}$ (DF4) at $m_{\rm vir} = 10^{11}\,{\rm M}_{\odot}$. Here, $x_{\rm bound} \equiv m_{\rm bound}/m_{\rm max,0}$ quantifies the strength of subhalo stripping and is jointly determined by the host halo potential, the satellite orbit, and the subhalo's internal structure. In cosmological simulations, the distribution of surviving subhalo mass fractions generally follows a log-normal form peaking near $\sim 10^{-1}$ \citep{Han2016, Nadler2023, He2023, He2026}, which implies that the differential probability decreases rapidly towards the heavy-stripping regime (very low $x_{\rm bound}$). Two additional considerations reinforce this preference against $m_{\rm vir} \sim 10^{11}\,{\rm M}_{\odot}$ progenitors: first, the cosmological subhalo mass function declines towards higher infall masses, making $10^{10.5}\,{\rm M}_{\odot}$ progenitors intrinsically more abundant; second, as shown in Sections~\ref{sec:selection} and~\ref{sec:implications}, cored, low-$n_{\rm sersic}$ satellites are highly susceptible to runaway disruption under extreme stripping, making survival down to $x_{\rm bound} \sim 10^{-3}$ particularly difficult. \textbf{Taken together, these effects indicate that the tidal progenitors of DF2 and DF4 are most likely to reside in cored haloes ($s_{001} \sim 0.3$) with masses near $m_{\rm vir} \sim 10^{10.5}\,{\rm M}_{\odot}$ and concentrations $c_{2} \sim 6\text{--}8$.}

Finally, we emphasise that our parameter scan only tests whether a track enters the target window. As discussed in the preceding section, the observable window for a DF2/DF4-like tidal remnant may span only a brief fraction of a satellite's orbital evolution, so evaluating the true detection likelihood must also account for the cosmological distribution of satellite orbits. We defer a detailed treatment of these orbital and survival-time effects to future work.

\section{Summary}\label{sec:summary}
We have used the tidal-track simulation suite of Paper~I to study how ultra-diffuse and DM-deficient satellite remnants form. The suite consists of 480 controlled $N$-body simulations of dwarf satellites. Each satellite starts with a DM halo of $m_{\rm vir}=10^{10}\,\Msun$ and orbits a static, Milky-Way-mass NFW host ($M_{\rm vir}=10^{12}\,\Msun$) with an initial apocentre of $100\,{\rm kpc}$ and a pericentre of 5--$50\,{\rm kpc}$. The initial conditions form a grid in halo concentration $c_2$, inner density slope $s_{001}$, stellar-to-halo mass ratio, stellar compactness $r_{\rm eff,0}/r_{\rm vir}$, and S\'ersic index $n_{\rm sersic}$ (Table~\ref{tab:grid}). DM and star particles have equal masses of $10^4\,\Msun$, and the stellar self-gravity is followed throughout stripping. In Paper~I, we summarised the evolution of the bound remnants into empirical tidal tracks for stellar mass, size, surface density, and velocity dispersion as functions of the bound mass, with track parameters that depend on the initial structure. In this paper, we have combined these tracks with the full simulations to characterise the conditions for forming DM-deficient ($m_{\star}/(m_{\star}+m_{\rm DM})(<r_{\rm eff})>1/2$) and ultra-diffuse ($r_{\rm eff}\geq1.5\,{\rm kpc}$ and $\langle\mu_{\rm e}\rangle_g\geq25\,{\rm mag\,arcsec^{-2}}$) remnants by tides. Our main results are as follows.
\begin{enumerate}[leftmargin=*]

\item We show that ultra-diffuse and DM-deficient morphologies generally correspond to opposing directions of tidal evolution. High-$n_{\rm sersic}$ stellar systems have enough stellar self-gravity to retain their stars and contract as their haloes are stripped, favouring compact, DM-deficient remnants. Conversely, low-$n_{\rm sersic}$ systems in cored subhaloes are more readily puffed up and stripped, yet require DM support to remain bound, favouring diffuse, DM-dominated remnants. Simultaneously producing both properties therefore requires a narrow transition regime between these two limiting pathways.

\item Consequently, objects that are simultaneously ultra-diffuse and DM-deficient, such as the observed NGC\,1052-DF2 and DF4, occupy this narrow transient regime. Only $4$ of our $480$ satellites, drawn from a broad, systematic grid of structural parameters, pass through the joint DMD-UDG region, and all four originate from cored haloes ($s_{001}=0.6$) with low S\'ersic indices ($n_{\rm sersic}\leq1$). Moreover, all four candidates survive in the DMD-UDG phase for only one or two apocentre passages before undergoing runaway disruption. Forming DF2/DF4-like galaxies through tidal stripping is therefore possible in principle, but it demands a restrictive combination of progenitor structure and stripping history and produces a short-lived transient phase rather than a stable endpoint.

\item We examine how the tidal tracks appear on scaling diagrams. In the heavily stripped phase, DM-deficient dwarfs can move approximately parallel to the classical Faber--Jackson reference on the $m_{\star}$--$\sigma_{\rm los}$ plane, while UDGs remain offset because of their larger DM contribution. On the $r_{\rm eff}$--$\Sigma_{\rm eff}$ plane, the tidal direction is set mainly by $\beta$: low-$\beta$ systems move towards lower surface density, while high-$\beta$ systems contract towards higher surface density.

\item Combining our empirical tidal-track model with cosmological priors on progenitor structure across $m_{\rm vir} = 10^{9.5}$--$10^{11}\,\Msun$ (Section~\ref{sec:progenitors}), we find that cuspy haloes are ruled out as viable tidal progenitors for DF2 and DF4 ($s_{001} \lesssim 1$ is strongly favoured). Lower-mass haloes ($m_{\rm vir} \lesssim 10^{10}\,\Msun$) are heavily suppressed by the SHMR prior because hosting $m_{\star} \sim 10^{8}\,\Msun$ lies well above their median stellar mass, whereas more massive haloes ($m_{\rm vir} \sim 10^{11}\,\Msun$) are less abundant and require substantially heavier tidal stripping ($x_{\rm bound} \sim 10^{-3}$--$10^{-2.8}$) that is increasingly rare in cosmological subhalo populations and more susceptible to complete disruption. Balancing these competing effects, our model favours cored progenitor haloes ($s_{001} \sim 0.3$) with masses around $m_{\rm vir} \sim 10^{10.5}\,\Msun$, concentrations $c_{2} \sim 6$--$8$, and initial stellar masses peaking near $m_{\star,0} \simeq 3 \times 10^{8}\,\Msun$.


\end{enumerate}

\begin{acknowledgments}
This work is supported by the National Natural Science Foundation of China (NSFC, Grant No. 12473007) and the China Manned Space Program (Grant No. CMS-CSST-2025-A03). The authors thank Xiaolong Du, Jiaxin Han, and Yimeng Tang for helpful discussions. F.J. thanks Aaron Ludlow for helpful discussions. F.H. thanks Zhaozhou Li for suggestions on constructing the initial conditions, and Guangze Sun for setting up and maintaining the Milkomeda computing server used for the simulations.
\end{acknowledgments}
\software{Agama \citep{Vasiliev2019}, Gadget-4 \citep{gadget4}, HBT+ \citep{Han2012,Han2018}, NumPy (\url{https://numpy.org}), SciPy (\url{https://scipy.org}), Astropy (\url{https://www.astropy.org}), Matplotlib (\url{https://matplotlib.org}).}

\section*{Data Availability}

The simulation data underlying this study are available
from the corresponding author upon reasonable request.


\bibliographystyle{aasjournalv7}
\bibliography{references}
\end{document}